\documentclass[useAMS,usegraphicx,usenatbib]{mn2e} 

\usepackage{graphicx}
\usepackage{times} 
\usepackage{amssymb}
\usepackage{amsmath}
\usepackage{lscape}
\usepackage{url}
\newif\ifAMStwofonts
\AMStwofontstrue

\def\suzaku{{\it Suzaku}}
\def\chandra{{\it Chandra}}

\def\xmm{{\it XMM-Newton}}

\def\spitzer{{\it Spitzer}}
\def\xspec{\hbox{\it XSPEC}}
\def\xspecv{{\it XSPEC}{\rm\thinspace v\thinspace 11.3.2}}

\def\frii{\hbox{\rm FR\thinspace II}}

\def\ks{\hbox{$\rm\thinspace ks$}}

\def\ghz{\hbox{$\rm\thinspace GHz$}}

\def\um{{\hbox{$\rm\thinspace \umu m$}}}

\def\kpc{\hbox{$\rm\thinspace kpc$}}
\def\mpc{\hbox{$\rm\thinspace Mpc$}}

\def\as{\hbox{$\rm\thinspace arcsec$}}

\def\pcmsq{\hbox{$\rm\thinspace cm^{-2}$}}
\def\kmpspmpc{\hbox{$\rm\thinspace km~s^{-1}~Mpc^{-1}$}}

\def\kev{\hbox{$\rm\thinspace keV$}}

\def\mjypb{\hbox{$\rm\thinspace mJy/beam$}}

\def\ctsps{\hbox{$\rm\thinspace count~s^{-1}$}}

\def\photpkevpcmsqps{\hbox{$\rm\thinspace ct~keV^{-1}~cm^{-2}~s^{-1}$}}
\def\photpcmsqps{\hbox{$\rm\thinspace ct~cm^{-2}~s^{-1}$}}

\def\ergpcmsqps{\hbox{$\rm\thinspace erg~cm^{-2}~s^{-1}$}}

\def\ergps{\hbox{$\rm\thinspace erg~s^{-1}$}}

\def\msun{\hbox{$\rm\thinspace M_{\odot}$}}

\def\qc{\hbox{\rm 6C\,0905+39}}

\begin{document}

\title[] {The Compton-thick quasar at the heart of the high-redshift
  giant radio galaxy \qc} \author[M. C. Erlund et al.]
{\parbox[]{6.in} {M.~C.  Erlund,$^{1}$\thanks{E-mail:
      mce@ast.cam.ac.uk} A.~C.
    Fabian$^{1}$, Katherine~M. Blundell$^{2}$\\ and Carolin~S. Crawford$^{1}$.} \\\\
  \footnotesize
  $^{1}$Institute of Astronomy, Madingley Road, Cambridge CB3 0HA\\
  $^{2}$University of Oxford, Astrophysics, Keble Road, Oxford OX1
  3RH\\ }
\maketitle

\begin{abstract}
  Our \xmm\ spectrum of the giant, high-redshift ($z=1.88$) radio
  galaxy \qc\ shows that it contains one of the most powerful,
  high-redshift, Compton-thick quasars known.  Its spectrum is very
  hard above 2\kev.  The steep \xmm\ spectrum below that energy is
  shown to be due to extended emission from the radio bridge using
  \chandra\ data. The nucleus of \qc\ has a column density of
  $3.5^{+1.4}_{-0.4}\times 10^{24}$\pcmsq\ and absorption-corrected
  X-ray luminosity of $1.7^{+0.9}_{-0.1}\times 10^{45}$\ergps\ in the
  $2-10$\kev\ band.  A lower redshift active galaxy in the same field,
  SDSS J090808.36$+$394313.6, may also be Compton-thick.
\end{abstract}


\begin{figure*}
\rotatebox{0}{
\resizebox{!}{8cm}
{\includegraphics{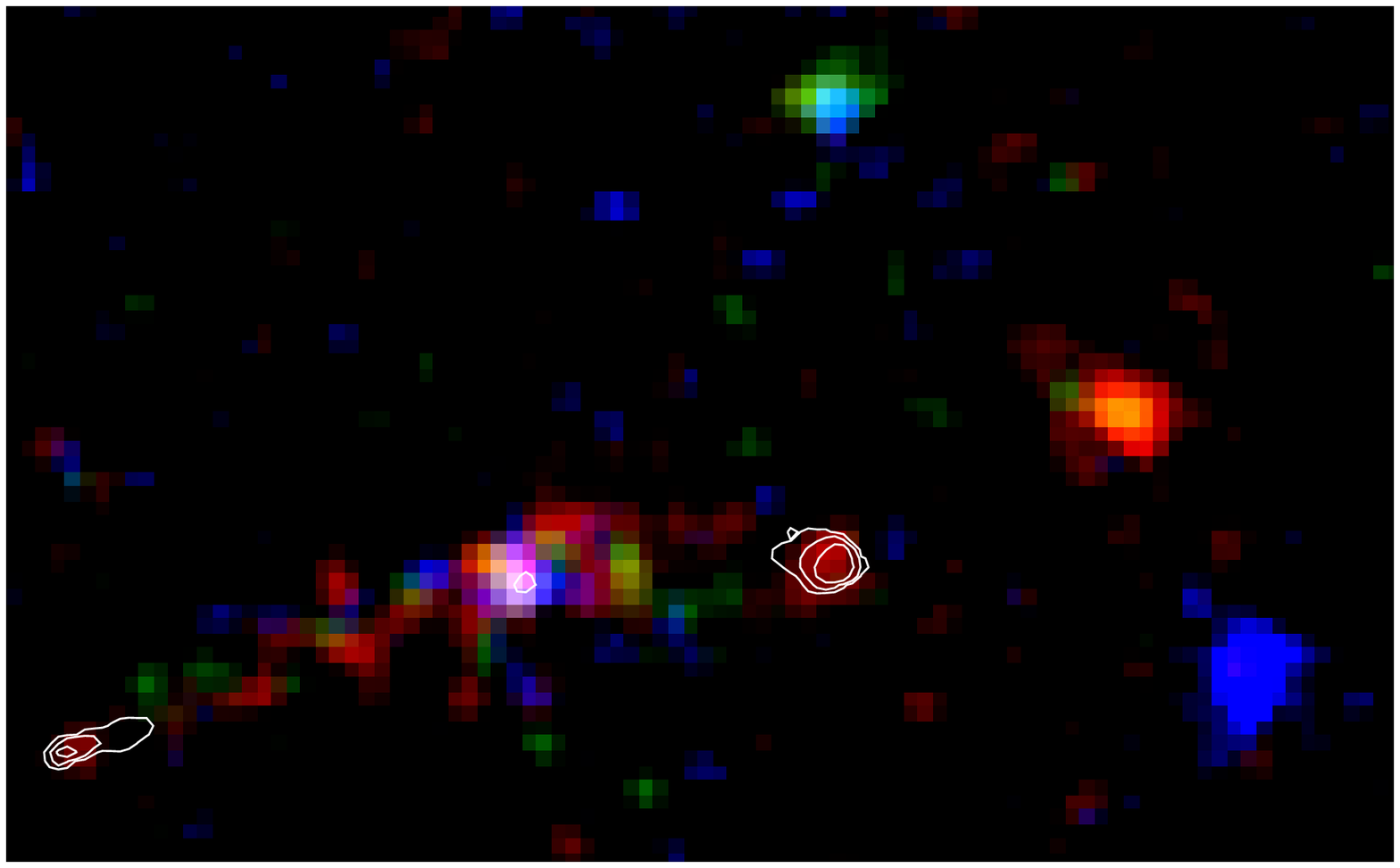}}}
\caption{Three colour plot of \xmm\ data ($188''\times 116''$ image
  with $2$\as\ pixels; the image has been smoothed by a Gaussian with
  $\sigma$ of $2$ pixels): $0.2 - 1$\kev\ band (red), $1-3$\kev\ band
  (green) and $3-10$\kev\ band (blue) all on a square-root scale. The
  white overlaid contours are VLA A-array $1.425$-\ghz\ radio contours
  ($0.4, 1.7$ and $7$\mjypb), displaying the core and two lobes of the
  radio source. The blue source to the west of \qc\ is an obscured AGN
  (see section \ref{sec:another}).}
\label{fig:colour}
\end{figure*}

\section{Introduction}

The powerful, classical double \frii\ \citep{FR} radio galaxy \qc\ is
unusual as it is one of the highest redshift ($z=1.88$) giant radio
galaxies known, spanning $111$\as\ on the sky \citep{lawgreen95}
corresponding to a projected size of $945$\kpc\ in the cosmology
assumed in this paper. Radio galaxies have long been associated with
obscured, (optically defined) Type 2 quasars, because they are
considered to be the same objects as radio-loud quasars seen through
an obscuring torus (e.g. \citealt{barthel89} and
\citealt{antonucci93}).  X-ray observations are particularly efficient
at detecting obscured AGN as they provide the observer with several
diagnostics, such as the photoelectric cut-off, the iron line
equivalent width (EW) and the X-ray to optical or [O III] $\lambda
5007$ ratio (e.g. \citealt{bassani99} or, for a review,
\citealt{comastri04}).  Sources with an obscuring column greater than
$1.5 \times 10^{24}$\pcmsq, at which the Thomson scattering optical
depth reaches unity, are known as Compton-thick AGN. It is non-trivial
to calculate the column density of such sources as there is little
(see \citealt{wilmanfabian99}) direct emission below 10\kev\ meaning
that analysis relies on interpreting the reflected spectrum at lower
X-ray energies. \suzaku\ is starting to provide a window into the
X-ray regime above 10\kev\ enabling direct detection of nuclear
emission in the brightest Compton-thick sources such as NGC\,5728
\citep{comastri07}. Such sources are important since the X-ray
background gives strong evidence for a significant population of
Compton-thick AGN (e.g. \citealt{comastri04}, \citealt{worsley05}).

In this paper, we present recent \xmm\ observations of \qc\ (which is
located at RA 09h08m16.9s, Dec +39d43m26s in J2000).  Throughout this
paper, all errors are quoted at $1\sigma$ unless otherwise stated and
the assumed cosmology is $\rm H_{\rm 0} = 71$\kmpspmpc, $\Omega_{0}=1$
and $\Lambda_{0} = 0.73$.  One arcsecond represents $8.518$\kpc\ on
the plane of the sky at the redshift ($z=1.883\pm 0.003$) of \qc\ and
the Galactic absorption along the line-of-sight is taken to be $1.91
\times 10^{20}$\pcmsq\ \citep{dickeylockman90}.


\section{Data Reduction}
\label{sec:reduction}

\xmm\ observed \qc\ on 2006 October 30 for $58.0$\ks\ with EPIC-pn and
$61.9$\ks\ with each of the EPIC-MOS cameras.  The data were reduced
using the standard pipeline, running the SAS tools EPCHAIN and EMCHAIN
for the EPIC-pn and both EPIC-MOS cameras respectively.  The resulting
files were filtered to remove flares, leaving $42.2$\ks\ of good-time
for the PN, and $53.1$\ks\ for both MOS\,1 and MOS\,2 when dead-time
intervals are also taken into account.

Spectra for the background (an area of sky free from sources near \qc\
on the same chip), the central source (within a $15$\as\ circle) and
the extended X-ray emission which is discussed elsewhere \citep{lobes}
were extracted separately for each instrument and stacked and fitted
using \xspecv. Given the low number of counts in the MOS spectra, only
the PN spectra were used for analysis of the core.

\section{Results}

The X-ray emission extending along the radio axis over hundreds of
\kpc\ to the east and west, which was first detected by \chandra\ and
presented in \citet{6c0905}, is clearly visible in the \xmm\ data (see
Fig.  \ref{fig:colour}). In total there are 2566 counts (1783
background) in the source, 2125 (1624 background) in the extended
emission and 441 counts (159 background) in the core in the
$0.2-8$\kev\ band after summing the MOS\,1, MOS\,2 and the PN data.
The small number of counts in the nucleus in our \chandra\ observation
\citep{6c0905} precluded any detailed analysis.  All spectral data
described in the present paper were fitted over the $0.5-10$\kev\
band.

\subsection{The extended emission}

The extended emission is discussed in detail in \citet{lobes};
however, we briefly mention it here in the context of its effect as a
potential contaminant of the core spectrum. Fig. \ref{fig:colour}
clearly shows the extended emission from the bridge of \qc\ and that
the number of X-ray photons increases towards the core. The extended
emission is not jet emission (since this would be very narrow and so
would not be resolved by \chandra), but inverse-Compton up-scattering
of the cosmic microwave background by spent radio plasma from the lobes \citep{6c0905}.

The bridge spectra (excluding a 15\as\ region around the core) were
extracted from the EPIC-pn, MOS\,1 and MOS\,2 data and were fitted
with a Galactic-absorbed power-law giving a photon index of
$\Gamma=1.75^{+0.27}_{-0.24}$ and a normalisation of
$2.5^{+0.3}_{-0.3}\times 10^{-6}$\photpkevpcmsqps\ with a
reduced-chisquared of $\chi^2_\nu = 0.9$ with $90$ degrees of freedom
(d.o.f.).  The absorption-corrected X-ray luminosity of the extended
emission in the $2-10$\kev\ band is $L_{\rm X} =
1.8^{+0.3}_{-0.3}\times 10^{44}$\ergps.

\subsection{The core}
\label{sec:core}

Only the PN data have been used in the following fits because each MOS
spectrum contains roughly a quarter of the counts in the PN spectrum
but a similar amount of noise. Spectral analysis shows that the MOS
spectra are consistent with the PN spectrum.

\begin{figure}
\rotatebox{0}{
\resizebox{!}{8cm}
{\includegraphics{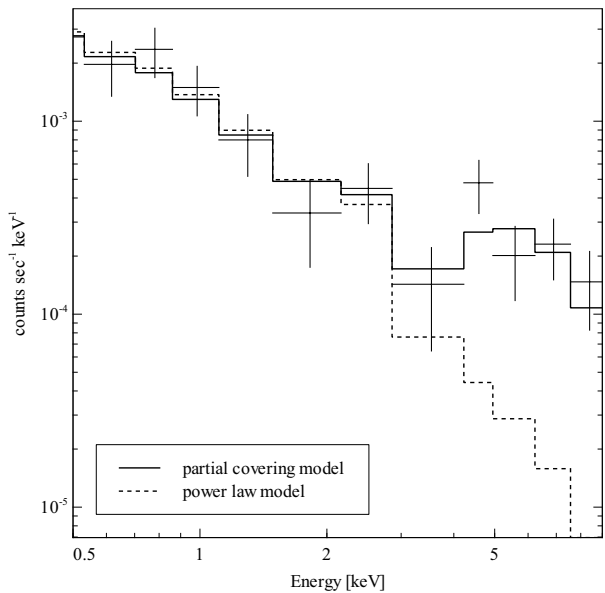}}
}
\caption{The spectrum of the core of \qc.  The solid line shows the
  best fit partial covering model and the dotted line shows the
  continuation of the directly observed power-law component.}
\label{fig:spec}
\end{figure}

The \xmm\ spectrum of the core is strikingly hard.  A simple Galactic
and intrinsic absorbed power-law is a poor fit with $\chi^2_\nu = 2.7$
for $8$ d.o.f. ($\chi^2 = 21.7$). The best-fit parameters give a
photon index of $\Gamma = 0.93$, a normalisation of $1.3\times
10^{-6}$\photpkevpcmsqps\ and no intrinsic absorption.  The spectrum
(Fig. \ref{fig:spec}) appears to have a power-law component with an
excess in the hard X-rays which can, in principle, be reproduced by
cold reflection (\xspec\ model $wabs$*$zwabs$*[$powerlaw$ +
$pexrav$]).  This model is not a good fit however giving $\chi^2_\nu =
2.2$ for $7$ d.o.f. ($\chi^2 = 15.4$) with an implausibly large amount
of reflection (reflection fraction $\sim 33$) and a photon index of
$\Gamma = 1.6$.

\begin{figure}
\rotatebox{90}{
\resizebox{!}{8.5cm}
{\includegraphics{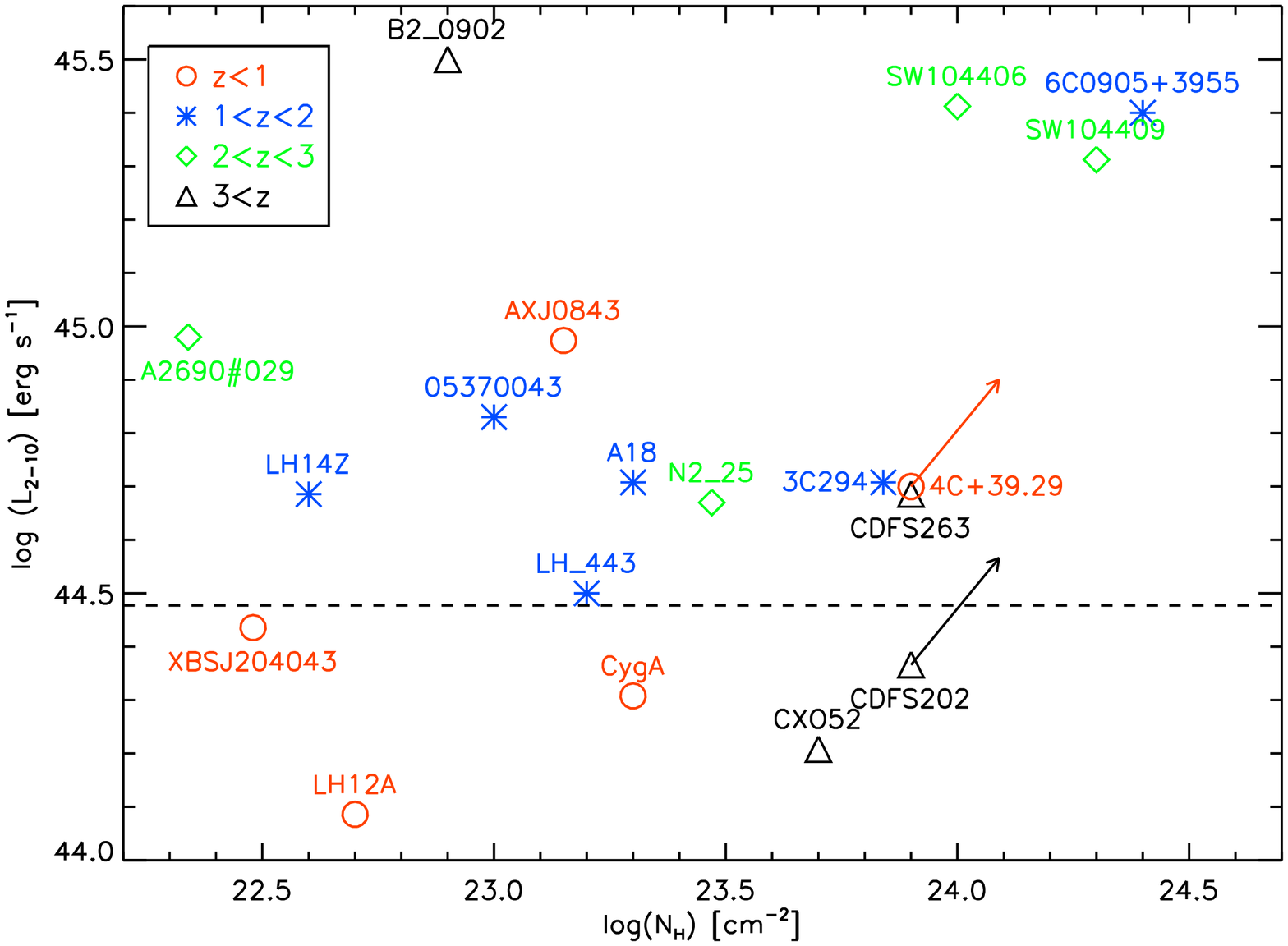}}
}
\caption{This figure is adapted from \citet{poshak} (their Fig. 7, see
  their caption for details of how this plot was produced and the
  relevant references for each source).  The X-ray luminosity axis
  shows the absorption-corrected luminosity in the $2-10$\kev\ band of
  each source.  B2 0902+39 (B2\_0902, \citealt{fabian02}), SWIRE
  J104406.30+583954.1 and SWIRE J104409.95+585224.8 (SW104406,
  SW104409 respectively; \citealt{polletta06}) and \qc\ have been
  added.  This selection of X-ray selected type 2 quasars from the
  literature excludes sources which are clearly reflection
  dominated. This implies that \qc\ is one of the most intrinsically
  luminous and one of the most obscured quasars yet detected.}
\label{fig:poshak}
\end{figure}

A partial-covering model with a redshifted iron line is a good fit
(Fig. \ref{fig:spec}). Including an iron line, the model gives a
covering fraction of $0.97^{+0.02}_{-0.03}$, intrinsic absorption of
$2.6^{+1.0}_{-0.7}\times 10^{24}$\pcmsq, photon index $\Gamma =
2.3^{+0.2}_{-0.3}$ and a normalisation of $5.0^{+0.7}_{-0.8}\times
10^{-5}$\photpkevpcmsqps\ for the power-law.  The redshifted iron line
(modelled as a Gaussian) was fixed at 6.4\kev\ in the rest-frame of
the source with $\sigma = 0.1$\kev, giving an equivalent width $EW
\sim 0.9$\kev; the normalisation of this component is
$2.2^{+1.5}_{-1.4}\times 10^{-5}$\photpcmsqps.  This gives $\chi^2_\nu
= 0.9$ with $6$ d.o.f. ($\chi^2 = 5.4$).  Without an iron line, the
fit values are similar: a covering fraction of $0.95^{+0.01}_{-0.04}$;
an intrinsic absorption of $2.1^{+1.0}_{-0.6}\times 10^{24}$\pcmsq; a
photon index $\Gamma = 2.1^{+0.2}_{-0.2}$ and a normalisation of
$3.2^{+0.6}_{-3.2}\times 10^{-5}$\photpkevpcmsqps, giving $\chi^2_\nu
= 1.1$ for $7$ d.o.f. ($\chi^2 = 7.8 $).  Including an iron line
improves the fit by $\Delta \chi^2 = 2.4$, so the data marginally
favour its addition. The absorption-corrected luminosity in the
$2-10$\kev\ band for the partial covering model with an iron line is
$L_{\rm X} = 3.2^{+0.3}_{-0.2} \times 10^{45}$\ergps\ and the observed
flux is $F_{\rm X} = 3\times 10^{-14}$\ergpcmsqps\ (the errors are
poorly constrained). The column density of \qc\ is that of a
Compton-thick AGN and its luminosity implies that it is a quasar. Such
Compton-thick quasars are rare, Fig. \ref{fig:poshak} (adapted from
\citealt{poshak}) illustrates this point.  It shows a selection of
X-ray selected type-2 quasars which are not reflection dominated. We
infer from this that \qc\ is particularly intrinsically luminous and
highly obscured.


The paucity of counts in the \chandra\ data ($14\pm 4$ in the
$0.5-7$\kev\ band) means that no analysis was performed on the
original detection of the core \citep{6c0905}.  The \chandra\ data are
consistent with the \xmm\ data in as much as they are incapable of
constraining the spectral shape of the nucleus. However, in the 2\as\
region extracted for the \chandra\ analysis there are significantly
fewer soft photons than in the \xmm\ data (15\as\ region).  To
quantify this we measure the observed count-rate below 1.5\kev, which
is $1.5\pm 0.9 \times 10^{-4}$\ctsps, and compare it to that predicted
from a model of the \xmm\ data in the same band, which gives a
count-rate of $5\times 10^{-4}$\ctsps.  This deficit is significant at
$\sim 4\sigma$ level and implies that the soft photons are from the
extended emission which is less of a contaminant in the smaller
\chandra\ aperture.

\begin{figure}
\rotatebox{0}{
\resizebox{!}{8cm}
{\includegraphics{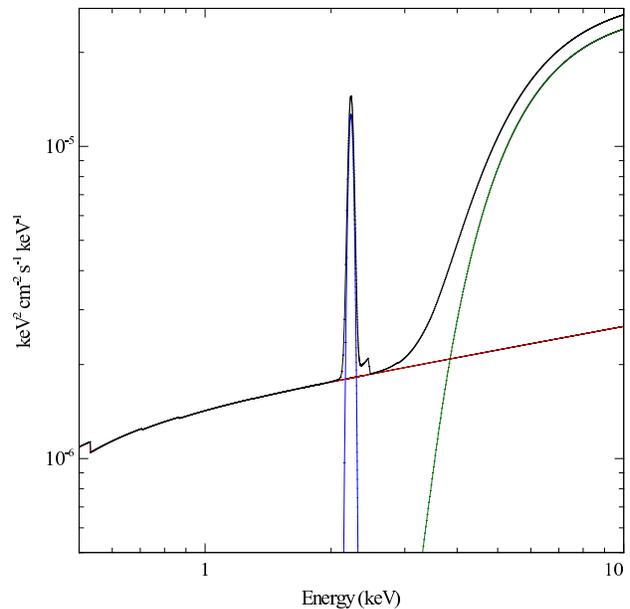}}
}
\caption{The best fit obscured AGN with contamination is shown in
  black. It consists of an absorbed Galactic and intrinsic absorbed
  power-law (green line); an additional power-law component fixed to
  have the same photon index as the extended emission ($\Gamma =
  1.75$, red line) and a redshifted iron line (blue line).}
\label{fig:contamination}
\end{figure}

Given that there is a considerable amount of extended emission from
the lobes of the radio source (clearly seen in Fig. \ref{fig:colour})
that is likely to be contaminating the \xmm\ nucleus spectrum, we now
try a highly obscured AGN with a power-law component at low energies
from the extended emission. In order to test this, we fit a model
consisting of Galactic absorption; a power-law with a photon index
fixed to the value found for the extended emission ($\Gamma = 1.75$,
see \citealt{lobes}); an intrinsically absorbed power-law (with a
photon index fixed to $\Gamma = 2$ as the data are unable to constrain
it); and a redshifted iron line with it energy fixed at $6.4$\kev\ and
$\sigma = 0.1$ (\xspec\ model $wabs[powerlaw + zwabs( powerlaw +
zgauss)]$).  The best-fitting model is shown in
Fig. \ref{fig:contamination}.  The intrinsic column density is
$3.5^{+1.4}_{-0.4}\times 10^{24}$\pcmsq\ and the normalisation of this
component is $2.7^{+1.4}_{-0.1}\times 10^{-5}$\photpkevpcmsqps.  The
redshifted iron line has a normalisation of $2.7^{+11.2}_{-2.5}\times
10^{-4}$\photpcmsqps. This model gives a good fit to the data with
$\chi^2_\nu = 1.1$ for $7$ d.o.f. ($\chi^2 = 7.4$); statistically the
partial covering model is a slightly better fit with $\chi^2 = 5.4$
for $6$ d.o.f. which gives $\Delta\chi^2 = 2$.  The observed flux is
$F_{\rm X} = 2.8^{+1.4}_{-0.5}\times 10^{-14}$\ergpcmsqps.  The
absorption-corrected luminosity in the $2-10$\kev\ band of the central
source is $L_{\rm X,core} = 1.7^{+0.9}_{-0.1}\times 10^{45}$\ergps.
The normalisation of the contaminating power-law is
$1.5^{+0.2}_{-0.2}\times 10^{-6}$\photpkevpcmsqps\ and its
absorption-corrected X-ray luminosity in the $2-10$\kev\ band is
$L_{\rm X,ext} = 1.1^{+0.1}_{-0.2}\times 10^{44}$\ergps.  The
luminosity of the extended X-ray emission is $L_{\rm X} =
1.9^{+0.3}_{-0.3}\times 10^{44}$\ergps\ implying that roughly a third
of the power in the emission due to inverse-Compton scattering of the
cosmic microwave background comes from close to the core.


\subsection{Another obscured AGN in the field?}
\label{sec:another}

\begin{figure}
\rotatebox{-90}{
\resizebox{!}{8cm}
{\includegraphics{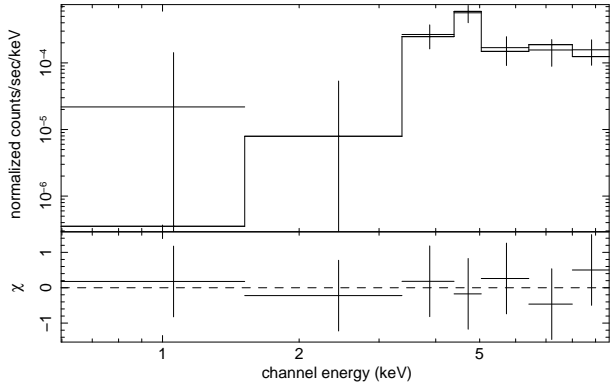}}
}
\caption{The PN spectrum of the second obscured AGN in the field (the
  blue source to the west of \qc\ in Fig. \ref{fig:colour}). The solid
  line is the best fit absorbed power-law model with a redshifted iron
  line.}
\label{fig:another}
\end{figure}

The hard X-ray source to the west of \qc\ (the blue source in
Fig. \ref{fig:colour} located at RA 09h08m08.3s, Dec +39d43m12s in
J2000 coordinates) is an SDSS source and so, following the SDSS naming
convention, is SDSS J090808.36$+$394313.6. It has a spectrum
consistent with it being an obscured AGN (see Fig. \ref{fig:another})
with column density of $N_{\rm H} \sim 1.28^{+0.41}_{-0.36}\times
10^{24}$\pcmsq\ when the power-law component is fixed at $\Gamma = 2$
because it is otherwise unconstrained (although varying it between
reasonable values does not significantly alter the inferred value of
the column density).  Adding an iron line redshifted from 6.4\kev\
improves the fit from $1.4$ for $4$ d.o.f. to $0.2$ for $3$ d.o.f.,
i.e. by $\Delta\chi^2 = 5$. The redshift is measured as $z =
0.45^{+0.02}_{-0.08}$. There are a total of 261 counts (159
background) in the $0.2-8$\kev\ band, meaning that the derived
quantities are somewhat uncertain; however, as Fig. \ref{fig:colour}
shows there are very few photons below 3\kev, this source must be a
highly obscured AGN. The $2-10$\kev\ band observed (absorbed) flux is
$F_{\rm X} = 3.1^{+5.7}_{-1.0}\times 10^{-14}$\ergpcmsqps\ and the
absorption-corrected X-ray luminosity in the same band is $L_{\rm X}
=1.3^{+0.3}_{-0.2} \times 10^{44}$\ergps.

\section{discussion}
\label{sec:discussion}

The spectral fitting of the nucleus of \qc\ presented in Section
\ref{sec:core} shows that a model consisting of an obscured quasar
nucleus with a contaminating power-law from the lobes is a good fit to
the data. Its column density and luminosity clearly means that this is
a Compton-thick quasar.

\citet{seymour07} have performed a \spitzer\ survey of 69
high-redshift radio galaxies and find that the monochromatic
luminosity of \qc\ at 8\um\ is $L_{\rm 8\mu m} \sim 5 \times
10^{44}$\ergps.  They do not have mid-infrared data for \qc, but in
the majority of cases where they have longer wavelength data, the
spectral energy distributions (SED) of their high-redshift radio
galaxies continue to rise, often by one to two orders of magnitude.
Observing \qc\ at these wavelengths is important to confirm its nature
as a Compton-thick quasar.

Radio galaxies have long been supposed to be radio quasars oriented
closer to the plane of the sky.  Given its immense linear size of
nearly 1\mpc, \qc\ is likely to lie close to the plane of the sky and
so, following the unification model, we are seeing the core through
the torus and thus we would expect the source to be Compton-thick.
This also means that beaming is unimportant.

The measured obscuring column and absorption corrected luminosity
suggests that \qc\ is a genuine Compton-thick quasar. (We note that
the column density measured directly from X-ray spectra depends on the
geometry of the obscuring medium as shown by \citealt{matt99}.
However, in the present case a simple torus geometry seems relevant.)
There are very few Compton-thick quasars known at high redshift.
\citet{norman02} detected a good Compton-thick quasar candidate in the
\chandra\ deep field South with $z>3$.  Apart from such clear
examples, very few Compton-thick quasars are known especially at
redshift $z\sim 2$, despite the fact that obscured AGN are expected to
make up $\sim 30$ per cent of the hard X-ray background and $70$ per
cent should be found between $1 < z < 3$, according to populations
synthesis models (see e.g. \citealt{norman02} and \citealt{gilli01}).

The absorption-corrected luminosity in the $2-10$\kev\ band can be
converted to the bolometric luminosity using results derived by
\citet{vasudevan07} who found that low and high Eddington
ratio\footnote{where the Eddington ratio is defined as the bolometric
  luminosity divided by the Eddington luminosity} AGN have a
bolometric correction of $19\pm 4$ and $55\pm 13$ respectively.  In
order to determine whether \qc\ is a low- or high-Eddington source,
the black hole mass is estimated using the black-hole-mass--bulge-mass
relationship \citep{magorrian98} and the stellar mass (measured by
\citealt{seymour07}) as a proxy for the bulge mass.  This gives a
black hole mass for \qc\ of $M_{\rm BH} \sim 2\times 10^{9}$\msun.
This is consistent with the value calculated using the method
presented in \citet{marconihunt03}, who make use of the
black-hole-mass--bulge-luminosity relation, and the K-band magnitude
(from \citealt{seymour07}) along with a crude redshift
correction. This assumes that such relationships hold to out to
redshift $z=1.88$ which is not necessarily true, but radio galaxies at
lower redshifts tend to be the hosts of similarly massive black holes
(e.g. \citealt{liu06}) so this estimate seems reasonable.  Using these
estimates for black hole mass, \qc\ appears to fall into the
high-Eddington regime ($L_{\rm bol}/L_{\rm Edd} > 0.1$) so the
appropriate bolometric correction is $55\pm 13$ which implies that the
bolometric luminosity of \qc\ is $L_{\rm bol} = 1.0\pm 0.4\times
10^{47}$\ergps.  This means that it is a very powerful
source. Compared to other powerful, obscured quasars (see
Fig. \ref{fig:poshak}, assuming that the measured column density is
accurate), \qc\ is amongst the most powerful, (X-ray defined) type 2
quasars currently known.

Recent research has shown that high-Eddington objects tend not to be
jetted (e.g. \citealt{panessa07}; \citealt{sikora07} and
\citealt{maoz07}). However, \qc\ does not seem to be follow this trend
as it is an extremely powerful source which is host to a radio source
which spans $945$\kpc\ on the sky and appears to still be actively
fed. Its giant size indicates that it has been continuously active for
at least the past ten million years.  Such high-redshift giant radio
galaxies may act as beacons for extremely powerful AGN.


\section{conclusions}
\label{sec:conclude}

Our \xmm\ spectrum of \qc, which is one of the highest redshift giant
radio galaxies known, shows it to contain a Compton-thick quasar.
\qc\ is one of very few such sources currently known and one of even
fewer at high redshift.  It is likely to be a high-Eddington source
with a bolometric luminosity of $\sim 10^{47}$\ergps, making it one of
the most powerful sources known.


\section*{Acknowledgements}

MCE acknowledges PPARC for financial support. ACF and KMB thank the
Royal Society. MCE thanks Poshak Gandhi for providing the data to
recreate his figure (Fig. \ref{fig:poshak}) so that \qc\ could be
included and Ranjan Vasudevan for helpful discussions.

\bibliographystyle{mnras} 
\bibliography{mn-jour,6c0905_xmm}
\end{document}